\documentclass[11pt]{article} %
\usepackage{graphicx}
\usepackage{fancyhdr}
\usepackage{makeidx}
\usepackage{amssymb,amsmath}
\usepackage{physics}
\usepackage{gensymb}
\usepackage{upquote}
\usepackage{ulem}
\usepackage{xcolor}
\usepackage{cite}
\usepackage{pdfpages}
\usepackage{chemgreek}
\usepackage[utf8]{inputenc}
\usepackage[nottoc]{tocbibind}
\usepackage{setspace}
\usepackage{enumerate}
\DeclareUnicodeCharacter{0308}{HERE!HERE!}
\newcommand{\bd}{\begin{document}}
	\newcommand{\ed}{\end{document}}
\newcommand{\bc}{\begin{center}}
	\newcommand{\ec}{\end{center}}
\newcommand{\vs}{\vspace}
\newcommand{\hs}{\hspace}
\newcommand{\beq}{\begin{equation}}
\newcommand{\eeq}{\end{equation}}
\newcommand{\beqs}{\begin{eqn*}}
	\newcommand{\eeqs}{\end{eqn*}}
\newcommand{\bq}{\begin{quote}}
	\newcommand{\eq}{\end{quote}}
\newcommand{\lb}{\linebreak}
\newcommand{\mb}{\makebox}
\newcommand{\fb}{\framebox}
\newcommand{\mc}{\multicolumn}
\newcommand{\ben}{\begin{enumerate}}
	\newcommand{\een}{\end{enumerate}}
\newcommand{\bit}{\begin{itemize}}
	\newcommand{\eit}{\end{itemize}}
\newcommand{\ov}{\overline}
\newcommand{\un}{\underline}
\newcommand{\lt}{\left}
\newcommand{\rt}{\right}
\newcommand{\ba}{\begin{array}}
	\newcommand{\ea}{\end{array}}
\newcommand{\beqa}{\begin{eqnarray}}
\newcommand{\eeqa}{\end{eqnarray}}
\newcommand{\beqas}{\begin{eqnarray*}}
	\newcommand{\eeqas}{\end{eqnarray*}}
\newcommand{\bfg}{\begin{figure}}
	\newcommand{\efg}{\end{figure}}
\newcommand{\pad}{\partial}
\newcommand{\nn}{\nonumber}
\newcommand{\la}{\leftarrow}
\newcommand{\ra}{\rightarrow}
\newcommand{\lgla}{\longleftarrow}
\newcommand{\lgra}{\longrightarrow}
\newcommand{\La}{\Leftarrow}
\newcommand{\Ra}{\Rightarrow}
\newcommand{\Lra}{\Leftrightarrow}
\newcommand{\Lgla}{\Longleftarrow}
\newcommand{\Lgra}{\Longrightarrow}
\renewcommand{\a}{\alpha}
\renewcommand{\b}{\beta}
\newcommand{\g}{\gamma}
\newcommand{\G}{\Gamma}
\renewcommand{\d}{\delta}
\newcommand{\D}{\Delta}
\newcommand{\e}{\epsilon}
\newcommand{\eps}{\epsilon}
\newcommand{\s}{\sigma}
\renewcommand{\l}{\lamda}
\newcommand{\m}{\mu}
\newcommand{\n}{\nu}
\renewcommand{\S}{\Sigma}
\newcommand{\p}{\pi}
\newcommand{\om}{\omega}
\newcommand{\Om}{\Omega}
\newcommand{\tri}{\triangle}
\newcommand{\ti}{\times}
\newcommand{\f}{\frac}
\newcommand{\ds}{\displaystyle}
\newcommand{\bm}[1]{\mb{{\boldmath $#1$}}}
\newcommand{\alter}[2]{\lt\{ \ba{ll}#1 \\ #2 \ea \rt.}
\newcommand{\alt}[4]{\lt\{ \ba{ll}#1 & \mb{if \, \,}#2 \\ #3 & \mb{}#4 \ea
	\rt.}
\newcommand{\altn}[4]{\lt\{ \ba{rl}#1 & \mb{if \, \,}#2 \\ #3 & \mb{}#4 \ea
	\rt.}
\newcommand{\altif}[4]{\lt\{ \ba{ll}#1 & \mb{if \, \,}#2 \\ #3 &
	\mb{if \, \,}#4 \ea \rt.}
\newcommand{\altnif}[4]{\lt\{ \ba{rl}#1 & \mb{if \, \,}#2 \\ #3 &
	\mb{if \, \,}#4 \ea \rt.}
\newcounter{algc}
\newcounter{romc}
\newcounter{Alphc}
\newcommand{\bl}{\begin{list}{{\it Step} ~\arabic{algc}~:} {\usecounter{algc}
			\setlength{\topsep}{0pt} \setlength{\itemsep}{0pt}}}
	\newcommand{\el}{\end{list}}
\newcommand{\blr}{\begin{list}{~\roman{romc}~:} {\usecounter{romc}
			\setlength{\topsep}{0pt} \setlength{\itemsep}{0pt}}}
	\newcommand{\elr}{\end{list}}
\newcommand{\bla}{\begin{list}{~\Alph{Alphc}~:} {\usecounter{Alphc}
			\setlength{\topsep}{0pt} \setlength{\itemsep}{0pt}}}
	\newcommand{\ela}{\end{list}}
\newcommand{\tsup}{\textsuperscript}
\newcommand{\tsub}{\textsubscript}

\newtheorem{theorem}{Theorem}
\begin{document}
	\title{Probing biexciton in monolayer WS\tsub2 through controlled many-body interaction}
	\author{Suman Chatterjee$^{1}$, Sarthak Das$^{1}$, Garima Gupta$^{1}$,\\Kenji Watanabe$^2$, Takashi Taniguchi$^3$ and Kausik Majumdar$^{1*}$\\
		$^1$Department of Electrical Communication Engineering, \\Indian Institute of Science, Bangalore 560012, India\\
		$^2$Research Center for Functional Materials,\\ National Institute for Materials Science, 1-1 Namiki, Tsukuba 305-044, Japan\\
		$^3$International Center for Materials Nanoarchitectonics,\\ National Institute for Materials Science, 1-1 Namiki, Tsukuba 305-044, Japan\\
		$^*$Corresponding author, email: kausikm@iisc.ac.in}
\maketitle
\begin{abstract}
 The monolayers of semiconducting transition metal dichalcogenides host strongly bound excitonic complexes and are an excellent platform for exploring many-body physics. Here we demonstrate a controlled kinetic manipulation of the five-particle excitonic complex, the charged biexciton, through a systematic dependence of the biexciton peak on excitation power, gate voltage, and temperature \textcolor {black}{using steady-state and time-resolved photoluminescence (PL). With the help of a} combination of the experimental data and a rate equation model, we argue that the binding energy of the charged biexciton is less than the spectral separation of its peak from the neutral exciton. We also \textcolor {black} {note} that while the momentum-direct radiative recombination of the neutral exciton is restricted within the light cone, such restriction is relaxed for a charged biexciton recombination due to the presence of near-parallel excited and final states in the momentum space.
  \end{abstract}
 \newpage
 \section*{Introduction:}
 Due to strong confinement and relatively large carrier effective mass, monolayers of semiconducting transition metal dichalcogenides (TMDCs) exhibit light-matter interaction strongly driven by excitonic complexes, and are thus an excellent platform for exploring many-body physics at the two-dimension \cite{chernikov2014exciton,berkelbach2013theory,ugeda2014giant,das2019layer}. Experimental Observations of neutral bright and dark excitons ($X^0$ and $X_D^{0}$) \cite{zhang2015experimental,chernikov2014exciton,ye2018efficient,das2019layer,zhang2017magnetic,he2014tightly}, and other excitonic complexes such as positive or negatively changed trion ($X^\pm$)  \cite{mak2013tightly,singh2016trion,kallatt2019interlayer,plechinger2016trion,das2020gate}, dark trion ($X^\pm_D$)\cite{li2019direct,zinkiewicz2020neutral,liu2019gate}, neutral biexciton ($XX^0$)\cite{ye2018efficient} and charged biexciton($XX^\pm$) \cite{you2015observation, ye2018efficient,barbone2018charge} have been widely reported. \textcolor{black}{One recent report also explained electroluminescence from exciton and its complexes\cite{paur2019electroluminescence}.} The charged biexciton is of particular interest, which is a five-particle complex, consisting of two holes and three electrons ($XX^-$) or three holes and two electrons ($XX^+$) bound through coulomb interaction, and are distinct from the neutral biexciton. Due to the specific formation and recombination kinetics, the rate equations dictate a square law dependence between the exciton and biexciton intensities\cite{asnin1972radiative,haynes1960experimental,pei2017excited,nagler2018zeeman,phillips1992biexciton}. This makes the biexciton the strongest emitting complex among all the observable excitonic complexes at higher excitation density \cite{pei2017excited,barbone2018charge,plechinger2015identification,sie2015intervalley}. Besides, the charged nature of the $XX^\pm$ species makes it amenable to be controlled by a gate voltage \cite{ye2018efficient,barbone2018charge}.\\
The identification of biexciton by superlinear power-law \cite{barbone2018charge,you2015observation,ye2018efficient,haynes1960experimental,pei2017excited}, its coherent formation from two excitons \cite{steinhoff2018biexciton}, and the binding energy \cite{kidd2016binding,kylanpaa2015binding} have been discussed in the recent literature. For tungsten-based TMDC, the neutral versus charged biexciton can be distinguished \cite{ye2018efficient,chen2018coulomb}. Interestingly, several of these works report a power law with exponent in the range of 1.4-1.6 for the charged biexciton \cite{ye2018efficient,chen2018coulomb,plechinger2015identification,pei2017excited}, which is significantly less than 2 - a physical explanation of the same will be desirable. Besides, a detailed understanding of the formation and recombination kinetics will be widely useful in the context of the biexciton. We address these questions in the present work.

 \section*{Results and Discussions:}
We probe the biexciton states using steady-state and time-resolved photoluminescence (PL) measurement. The lowest excitonic state in 1L-WS\tsub2 is spin-dark due to its unique spin-split bands at $K(K^\prime)$ valleys \cite{zhang2015experimental,zhang2017magnetic,echeverry2016splitting,robert2017fine}. Under the steady-state condition, as a result of a larger radiative lifetime of the dark exciton state \cite{tang2019long,zinkiewicz2020neutral,robert2017fine}, a large fraction of the light-emitting neutral biexcitons ($XX^0$) form through each pair of spin-bright and spin-dark excitonic states, as opposed to both being bright exciton \cite{li2019direct,ye2018efficient}. Further, in the presence of doping, the formation of a five-particle charged biexciton ($XX^\pm$) is favored. There likely exists a larger steady-state density of completely dark charged-biexciton (forming from a dark exciton and a dark trion) in WS\tsub2, though it does not contribute to the PL signal.
	
\begin{figure}[!hbt]
		\centering
		\vs{-0.1in}
		\hs{-0.0in}
		\includegraphics[scale=0.5]{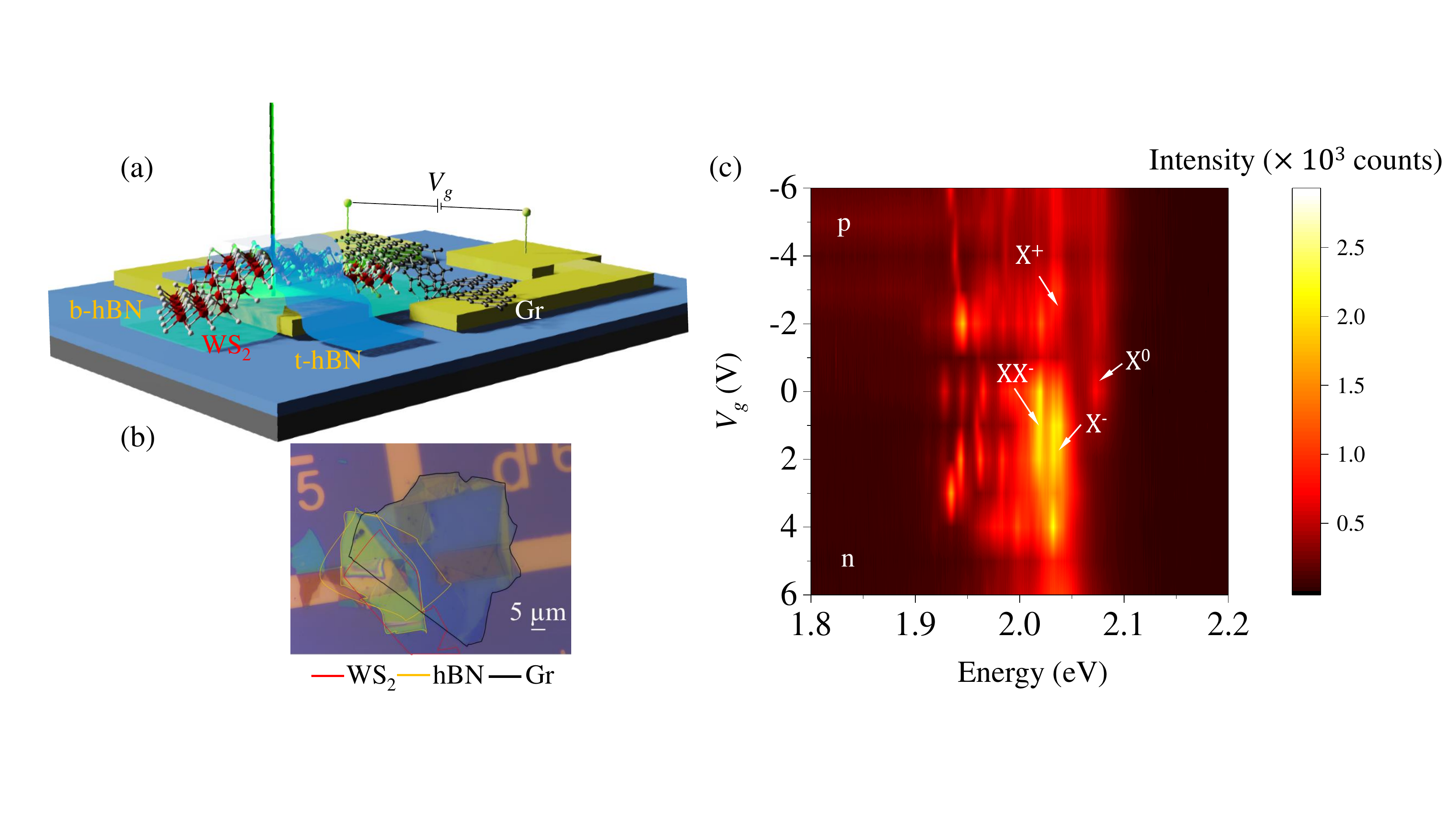}
		\vspace{-0.5in}
		\caption{\textbf{Device schematic and gate modulation of emission from excitonic complexes.}
				(a) Schematic and (b) Optical image of the device D1 (1L-WS\tsub{2} sandwiched between top and bottom hBN layers) fabricated for controlled electrical doping, a gate voltage ($V_g$) is applied to the bottom Au electrode while WS\tsub{2} is connected to another grounded electrode.
				(c) Doping dependent intensity variation of exciton ($X^0$), trion ($X^\pm$) and charged biexciton ($XX^-$) shown and \textcolor{black}{marked (by arrows)} in a color plot. The presence of charged biexciton ($XX^-$) is clearly associated with the presence of both the trion and exciton together.} \label{fig:gating}
	\end{figure}

	To identify the type of biexcitons ($XX^0$ or $XX^\pm$) in our experiment, we prepare an Au/hBN/1L-WS\tsub2/hBN (sample D1, represented in Figure \ref{fig:gating}a-b) using dry transfer method (see \textbf{Methods} for sample preparation). A gate voltage ($V_g$) is applied to the bottom Au electrode, while connecting the WS\tsub2 layer to a grounded few-layer graphene contact. The $V_g$ dependent PL spectra, taken at $4$ K, is shown in a color plot in Figure \ref{fig:gating}c (see \textbf{Supporting Information \textcolor{black}{Figure 1}} for representative spectra and \textcolor{black}{fitted peak intensity values}), suggesting a strong modulation of the neutral exciton ($X^0$), trion ($X^-$) and $XX^-$ peak intensity, in good agreement with previous reports \cite{barbone2018charge, plechinger2015identification,das2020highly,nagler2018zeeman}. With $V_g> 0$, the intensity of $X^0$ is suppressed, while that of $X^-$ is enhanced. \textcolor{black}{The $XX^-$ peak is observed in the $V_g$ range where both $X^0$ and $X^-$ have PL signatures, suggesting the origin of the five-particle state from the neutral and charged excitons. Note that the maximum intensity of the $XX^-$ occurs at a slightly higher positive $V_g$ compared to the maximum intensity of $X^0$. Such a reduction in the $X^0$ intensity at the maximum $XX^-$  intensity point results from a non-radiative loss of the excitons towards formation of trions and charged biexcitons}.
	
\begin{figure}[!hbt]
		\centering
		\vs{-0.1in}
		\hs{-0in}
		\includegraphics[scale=0.5]{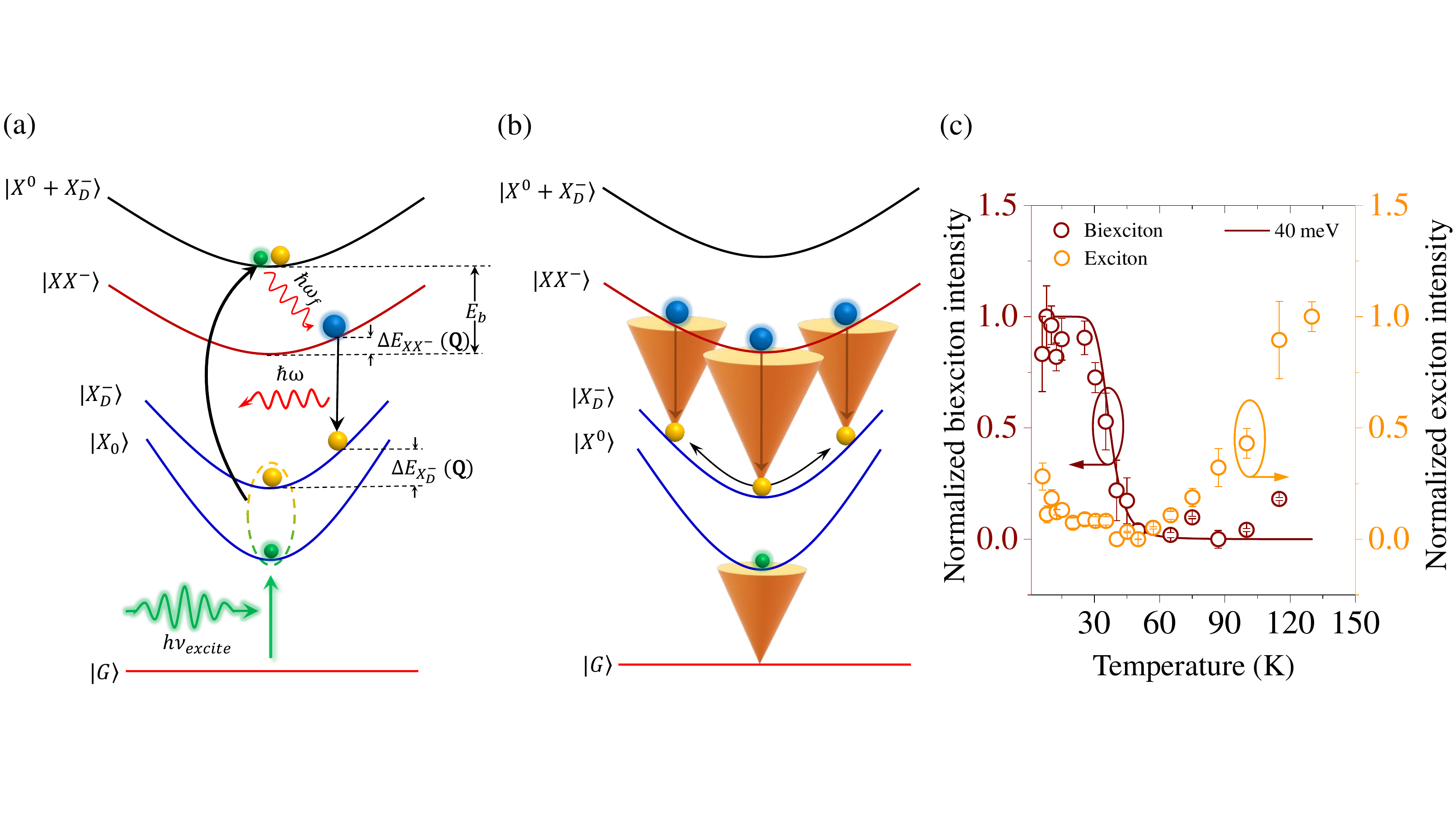}
		\vspace{-0.5in}
        \caption{\textbf{Model for formation, radiative recombination of charged biexciton and temperature dependent intensity of charged biexciton 1L-WS\tsub2}
        (a) Creation and recombination of charged biexciton ($XX^-$) is shown using parallel band model. Dark trion ($X_D^-$) and bright exciton ($X^0$) together form $XX^-$, with a binding energy $E_b$ and a kinetic energy $\Delta E_{XX^-}(Q)$ through the emission of a phonon of energy $\hbar\omega _f$. During  recombination, one photon is emitted of energy $\hbar \omega$, and the final state is a dark trion ($X_D^{-}$) with a recoil energy of $\Delta E_{X_D^-}(Q)$.
        (b) Multi-light-cone model for radiative recombination of the charged biexciton.
        (c) Normalized $X^0$ and $XX^-$ emission intensities plotted as a function of temperature, indicating opposite trends for $X^0$ and $XX^-$. The Fitted traces to the $XX^-$ peak intensity indicate a binding energy of \textcolor{black}{$40 \pm 5$ meV}.
    }\label{fig:bandstructure}
	\end{figure}

	Figure \ref{fig:bandstructure}a schematically depicts the formation process of the charged biexciton where a dark trion ($X_D^-$) \cite{zinkiewicz2020neutral,zinkiewicz2020excitonic} and a bright $X^0$ collide over a cross-section at the band edge, and form a $XX^-$ with kinetic energy $\Delta E_{XX^-}$, along with emitting a phonon of energy $\hbar\omega_f$. This leads to
	\begin{equation}\label{eq:form}
		E_{X^0}+E_{X_D^-}= E_{XX^-}(\mathbf{Q}=\mathbf{0})+\underbrace{\Delta E_{XX^-}(\mathbf{Q})+ \hbar\omega_f}_{E_b}
	\end{equation}
	Here, $E_{X^0}$, $E_{X_D^-}$ and $E_{XX^-}$ are the energies of the bright exciton, dark trion and charged biexciton, respectively. By this process the dark trion and the bright exciton bind together with a binding energy given by
	\begin{equation}\label{eq:Eb}
		E_{b}= \Delta E_{XX^-}(\mathbf{Q}) + \hbar\omega_f
	\end{equation}
	During recombination at finite $\mathbf{Q}$, there is a direct transition from the $XX^-$ band to the $X_D^-$ band, emitting a photon of energy $\hbar\omega$, leading to
	\begin{equation}\label{eq:recomb}
		E_{XX^-}(\mathbf{Q}=\mathbf{0})+\Delta E_{XX^-}(\mathbf{Q})= E_{X_D^-}(\mathbf{Q}=\mathbf{0})+\Delta E_{X_D^-}(\mathbf{Q})+ \hbar\omega
	\end{equation}
	where $\Delta E_{X_D^-}(\mathbf{Q})$ is the recoil energy of the remaining dark trion. There is a striking difference in the final states of the biexciton and the exciton recombination process. Linear momentum conservation forces all recombination giving out a photon to occur within the light cone of the exciton band around $\mathbf{Q}=\mathbf{0}$ \cite{gupta2019fundamental,yu2014dirac}. However, such constraint is relaxed for the biexciton recombination. Since the final state ($X_D^-$) is varying in energy with Q, the recombination process of $XX^-$ can also take place at varying Q, \textcolor{black}{as long as the final state is empty and the transition is allowed.} This can be visualized as multiple light cones, schematically depicted in Figure \ref{fig:bandstructure}b. \\
Combining Equations \ref{eq:form}, \ref{eq:Eb}, and \ref{eq:recomb}, we get
	\begin{equation}\label{eq:Eb1}
		E_{b}= (E_{X^0} - \hbar\omega) - [\Delta E_{X_D^-}(\mathbf{Q}) - \Delta E_{XX^-}(\mathbf{Q})]
	\end{equation}
	Note that $XX^-$ being a heavier particle than $X_D^-$, the bands are non-parallel with $\Delta E_{X_D^-}(\mathbf{Q}) - \Delta E_{XX^-}(\mathbf{Q}) \approx \frac{\hbar^2Q^2}{15m_0} > 0$ for non-zero $\mathbf{Q}$ (assuming $m_e = m_h \approx 0.5m_0$). Thus the binding energy of the biexciton is less than the separation between the exciton and the biexciton PL emission peaks ($E_{X^0}-\hbar\omega > E_b$), with the deviation being quadratic in $Q$.
	
	In order to establish this, in \textcolor{black}{Figure \ref{fig:bandstructure}c}, we show the temperature-dependent PL intensity variation of $X^0$ and $XX^-$, obtained from PL spectra from a Ag/hBN/1L-WS\tsub2/hBN stack (sample D2, see \textbf{Supporting Information Figure 2a and b} for device schematic and optical image) with no active gating and keeping WS\tsub2 electrically floating. The thickness of the bottom and top hBN layer is kept in the range of 25-30 nm and 10-15nm, respectively. The thickness is chosen to enhance the signal through constructive interference in the stack\cite{epstein2020near,dandu2020spectrally}. We observe the charged biexciton emission up to a sample temperature of 130 K. The Ag layer provides a pathway for fast dissipation of locally generated heat due to the focussed laser spot, thus helping to suppress laser-induced local heating. The separation between the exciton and charged biexciton peaks ($E_{X^0}-\hbar\omega$) is $60 \pm 5$ $meV$. The plot of intensity variation of $X^0$ peak as a function of $T$ in \textcolor{black}{Figure \ref{fig:bandstructure}c} shows an increasing trend with $T$. We attribute this to the fact that the bright exciton being energetically higher than the dark exciton state in W-based TMDCs \cite{zhang2015experimental,zhang2017magnetic,zinkiewicz2020neutral}. However, the $XX^-$ peak shows the opposite trend, with sharply decreasing peak intensity (see \textbf{Supporting Information \textcolor{black}{Figure 3a} and b} \textcolor{black}{for model spectrum fitting} and temperature-dependent spectra respectively) as temperature increases, in agreement with previous reports \cite{you2015observation,chen2018coulomb}. The data fits well with an equation of the form\textcolor{black}{\cite{chen2012photoluminescence, savenije2014thermally}} (see \textbf{Supporting Information S1}):
	\beq \label{eq:decay(T)}
	\eta(T)= \frac{1}{1+\kappa e^{-E_b/k_BT}}
	\eeq
	suggesting an enhanced thermal dissociation of $XX^-$ at higher temperature. From the fit, we estimate a value of $E_b$ in the range of \textcolor{black}{$40 \pm 5$ meV}, which is less (15 - 20 meV) than the peak separation in the PL spectra, and in agreement with our previous assertion $E_{X^0}-\hbar\omega > E_b$.
	

To understand the strong temperature dependence of the kinetics of the charged biexciton population, we write the rate equations using the formation of this five particle complex from its constituent particles\textcolor{black}{\cite{ye2018efficient,asnin1972radiative}}:
\begin{equation}
	dn_{X^0}/dt= -n_{X^0}/\tau_{X^0} + \beta_1P + \gamma_2 n_f n_{XX^-} e^{\frac{-\Delta E_{XX^-}}{k_BT}} - \beta_2 n_{X^0} - \gamma_1 (n_f+1)n_{X^0}n_{X^-_{D}}
\end{equation}
\begin{equation}
	dn_{X^-_{D}}/dt=-n_{X^-_{D}}/\tau_{X^-_{D}} + \beta_2 n_{X^0} + \gamma_2 n_f n_{XX^-} e^{\frac{-\Delta E_{XX^-}}{k_BT}} +  n_{XX^-}/\tau_{XX^-} - \gamma_1 (n_f+1)n_{X^0}n_{X^-_{D}}
\end{equation}
\begin{equation}
	dn_{XX^-}/dt=-n_{XX^-}/\tau_{XX^-} + \gamma_1(n_f+1)n_{X^0} n_{X^-_{D}} - \gamma_2 n_f n_{XX^-} e^{\frac{-\Delta E_{XX^-}}{k_BT}}
\end{equation}
 The parameters $\beta_1$,$\beta_2$, $\gamma_1$ and $\gamma_2$ are proportional to the quantum efficiency of creation of $X^0$ from incident photons, the rate of decay of $X^0$ to $X_D^-$ population, the rate of $XX^-$ creation and the rate of $XX^-$ dissociation, respectively.
$n_f$ is the phonon number. $n_{X^0}$, $n_{X^-_{D}}$  and $n_{XX^-}$ represent density of the respective excitonic species. $\tau_{X^0}$, $\tau_{X^-_{D}}$ and $\tau_{XX^-}$ represent radiative lifetime of bright exciton, dark trion and biexciton respectively.

The steady-state solution of the set of equations, if thermal-equilibrium phonon distribution (Bose-Einstein distribution) is considered, predicts $n_{XX^-} \propto n_{X^0}^2$, that is, a square-law dependence of biexciton population with respect to exciton population. However, with higher incident photon flux, every biexciton formation gives rise to the emission of a phonon. \textcolor{black}{Since phonons are relatively long-lived, $n_f$  deviates from equilibrium Bose-Einstein distribution.} \textcolor{black}{Under such a condition, we approximate $n_f \propto n_{XX^-}.$} This leads to a generation of the non-equilibrium hot phonon density, which in turn triggers an efficient dissociation of $XX^-$, forcing a deviation from quadratic law. In this condition, we obtain (see \textbf{Supporting Information S2} for detailed calculation):
\begin{equation}\label{eq:expl}
	An_{XX^-} + Bn_{XX^-}^2 exp(-\Delta E_{XX^-}/k_BT) = n_{X^0}^2
\end{equation}
with {$A = \frac{1}{\tau _{XX^-} \gamma _1 (n_f+1) \beta _2 \tau _{X_{D}^-}}$ and $B = \frac{\gamma _2\tau _{XX^-}}{\tau _{XX^-} \gamma _1 (n_f+1) \beta _2 \tau _{X_{D}^-}}$}
\\
This establishes a one-to-one correspondence between charged biexciton ($n_{XX^-}$) and bright exciton ($n_{X^0}$) population and clearly indicates a deviation from the near-equilibrium square law dependence at for $T>0$. Equation \ref{eq:expl} also suggests that the corresponding kinetics can be tuned by external stimulus. \textcolor{black}{Note that the primary temperature dependence arises from the factor $e^{\frac{-\Delta E_{XX^-}}{k_BT}}$, while other parameters are very weak function of temperature. Accordingly, with an increase in temperature, the second term on the left-hand side starts dominating, and the equation becomes linear.}

\begin{figure}[!hbt]
		\centering
		\vs{-0.1in}
		\hs{-0.5in}
		\includegraphics[scale=0.5]{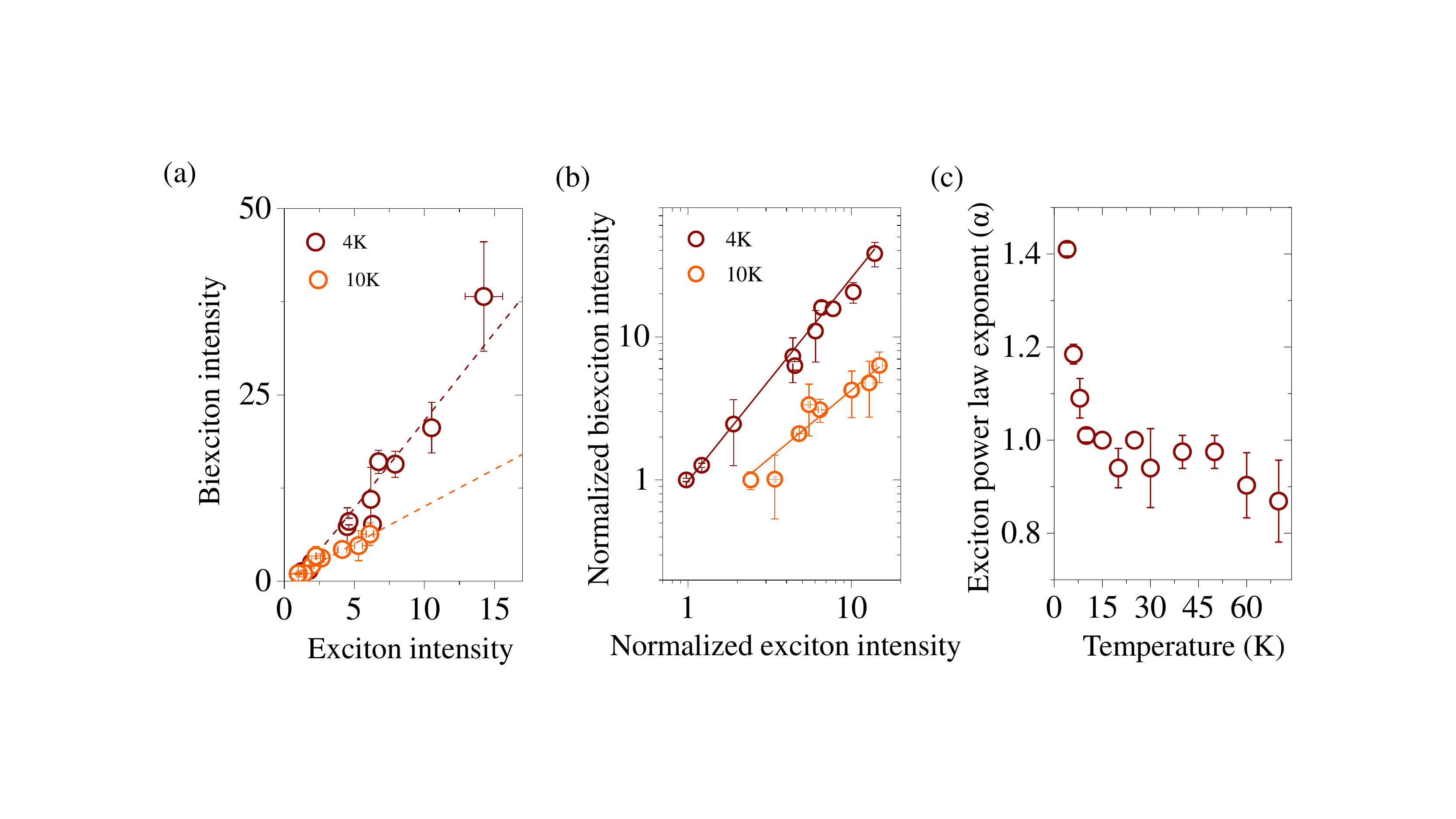}
		\vspace{-0.5in}
		\caption{\textbf{Temperature dependent kinetics for charged biexciton.}
			(a) Experimentally obtained intensity of $XX^-$ plotted as a function of the intensity $X^0$ species (symbols) at 4 K and 10 K. The dashed lines represent the fitted curves obtained from equation \ref{eq:expl}, plotted at different temperatures. (b) Log-log plot of $XX^-$ intensity with varying exciton ($X^0$) intensity. The \textcolor{black}{solid lines} indicate fits to the approximate power law: ($I_{XX^-} \propto I_{X^0}^{\alpha}$). (c) Plot of the $\alpha$ as a function of temperature, indicating a sharp fall with increasing temperature.}\label{fig:powerlaw}
	\end{figure}

To verify this temperature dependent coupled exciton - charged biexciton kinetics, we probe the excitation power ($P$) dependence of the PL spectra (on sample D2) from \textcolor{black}{$T=4.2$ to $70$ K} (representative spectra shown in \textbf{Supporting Information \textcolor{black}{Figure 4}}). Experimentally obtained $X^0$ and $XX^-$ PL intensities are fitted with equation \ref{eq:expl} using A and B as parameters, shown in Figure \ref{fig:powerlaw}a. The best fit is obtained at $\Delta E_{XX^-} \approx 10$ meV. It is evident from equation \ref{eq:expl} that a power law type relation such as $I_{XX^{-}}\propto I_{X^0}^\alpha$ does not exactly hold. However, the value of $\alpha$ from such a relation provides a quick qualitative understanding, and helps to compare with existing reports in literature. Figure \ref{fig:powerlaw}b represents such a power law fitting at 4 and 10 K (\textcolor{black}{Fittings at higher temperature values with equation \ref{eq:expl} and $I_{XX^{-}}\propto I_{X^0}^\alpha$ are shown in \textbf{Supporting Information Figure 5})}. The obtained $\alpha$ values are plotted as a function of temperature in Figure \ref{fig:powerlaw}c ($XX^-$ intensity variation with laser power ($P$) at different temperature points is fitted with $I_{XX^-} \propto P^\alpha$ and represented in \textbf{Supporting Information \textcolor{black}{Figure 6a and b}}).
We observe that $\alpha$  sharply drops with an increase in temperature, suggesting that the second term in the left-hand side of equation \ref{eq:expl} dominates the other term at a higher temperature \textcolor{black}{(This power law measurement is repeated for sample D1, which is the stack used for gating and fitted with $I_{XX^{-}}\propto I_{X^0}^\alpha$ lines. Obtained $\alpha$ values are shown in \textbf{Supporting Information Figure 10})}.
	
\begin{figure}[!hbt]
	\centering
	\vs{-0.1in}
	\hs{-0.5in}
	\includegraphics[scale=0.5]{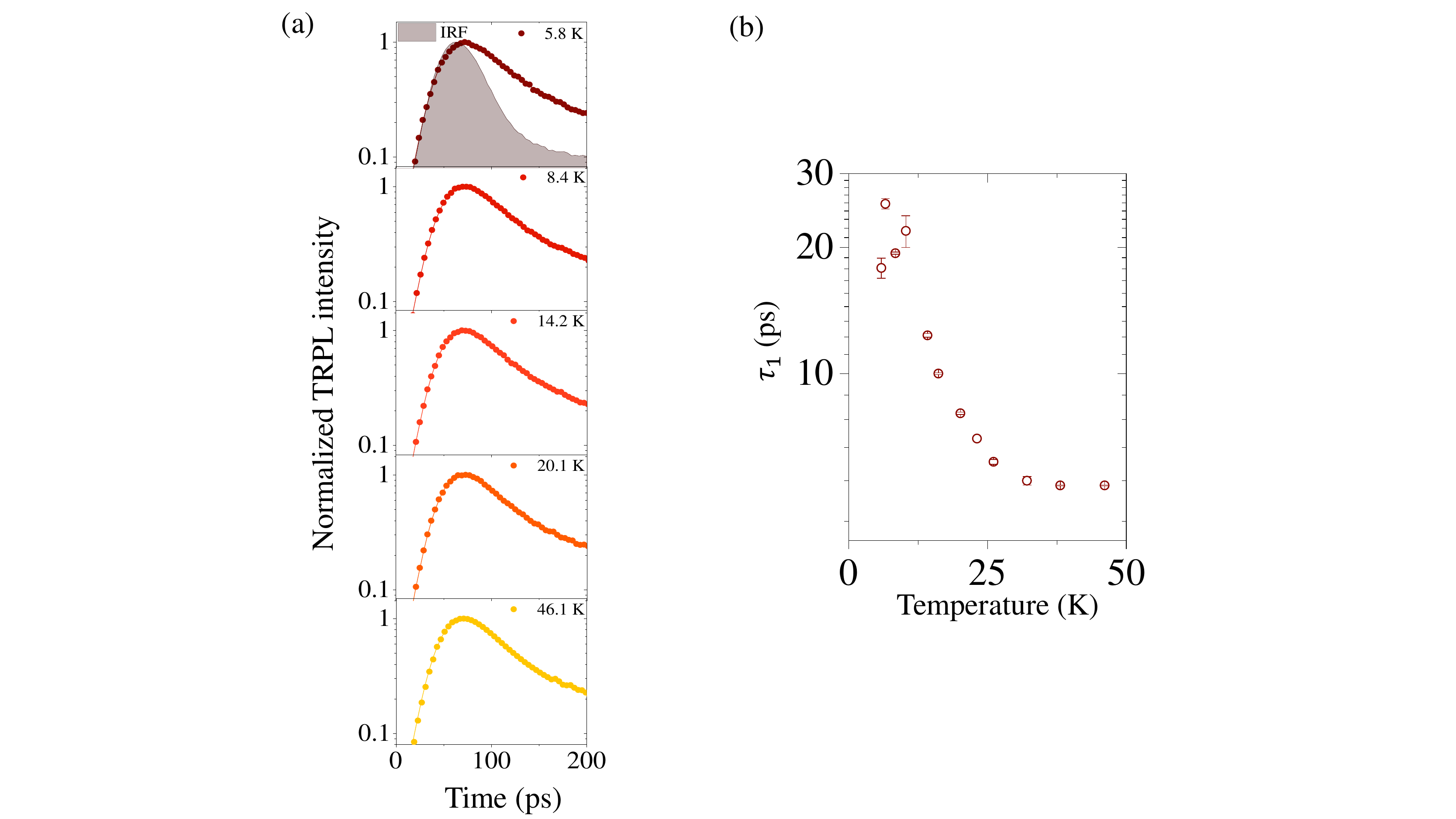}
	\vspace{-0.5in}
	\caption{\textbf{Temperature dependent lifetime of charged biexciton.}
		(a) TRPL with varying temperature from 5.8 K to 46.1 K. The symbols represent the measured data, and the solid traces represent the fitted curves (see \textbf{Methods}).
		(b) The decay time plotted as a function of temperature, as obtained from (a).}\label{fig:TRPL}
\end{figure}

To further investigate the temperature dependence of the $XX^-$ kinetics, we perform time-resolved photoluminescence (TRPL) measurement of 1L WS\tsub2 at different $T$. We excite the sample with a 531 nm pulsed laser (\textcolor{black}{Average optical power of 17 $\mu W$}) with an FWHM of 48 ps and a repetition rate of 5 MHz. The instrument response function (IRF) shows a full-width-at-half-maximum (FWHM) of 52 ps and a decay of 23 ps (See \textbf{Methods} for details about measurement technique). To filter out $XX^-$ luminescence only, we use a bandpass filter (10 nm passband) centered at 610 nm. \textcolor{black}{The \textit{in situ} steady-state PL spectra obtained at different temperatures are shown in \textbf{Supporting Information Figure 8}, with the filter window marked.} The time-dependent emission from the $XX^-$ peak at different temperatures is shown (in symbols) in \textcolor{black}{Figure \ref{fig:TRPL}a. To obtain the decay time of $XX^-$, we fit (solid lines) the measured TRPL data using two exponentials ($\sum_{i=1}^{2} A_i e^{-t/ \tau_i}$) after deconvoluting from the IRF as implemented in QuCoa software (PicoQuant). We note that the faster (and stronger) component ($\tau_1$) that represents the charged biexciton decay is strongly dependent on $T$. In Figure \ref{fig:TRPL}b we plot it as a function of $T$, which indicates that the net lifetime of the $XX^-$ becomes shorter with increasing $T$. In particular, we estimate $\tau_1 \approx 25 \pm 0.7$ ps at 6.6 K \cite{nagler2018zeeman,chowdhury2018ultrafast} and $\approx 5 \pm 0.13$ ps at 32 K}. Such a reduction in the lifetime is in excellent agreement with equation \ref{eq:expl} and the results in Figure \ref{fig:powerlaw}c, clearly suggesting a fast dissociation of $XX^-$ as temperature increases. It is instructive to consider the situation as follows: since both $X_D^-$ and bright $X^0$ are present in the system in a large number, the formation of $XX^-$ is not hindered. However, once the rapid dissociation starts at a higher temperature, it provides a fast non-radiative channel giving rise to reduced intensity, small lifetime, and near-unity power law. \textcolor{black}{We also note that there exists a weak (with relative weightage of $A_1/A_2 \approx 10 - 100$), slower decay component with $\tau_2 \sim 50-100$ ps. We believe this results from the tail of defect-bound excitonic complexes (see \textbf{Supporting Information Figure 8}), usually observed in WS$_2$.}

\begin{figure}[!hbt]
	\centering
	\vs{-0.1in}
	\hs{-0.5in}
	\includegraphics[scale=0.5]{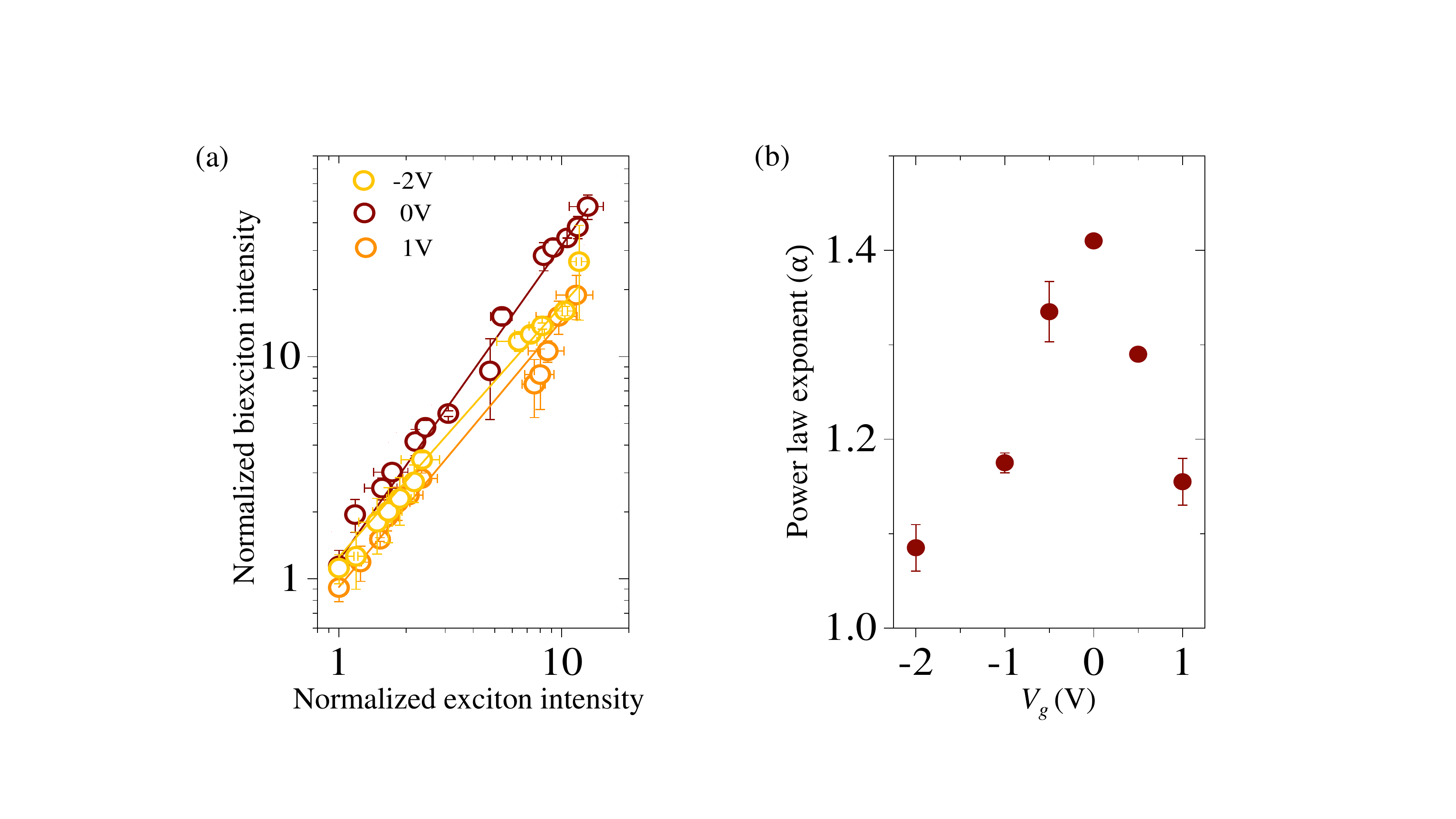}
	\vspace{-0.5in}
	\caption{\textbf{Gate voltage dependent charged biexciton kinetics.}
.
	(a) Plot of $XX^-$ peak intensity with varying exciton intensity ($I_{X_0}$) at different $V_g$. The \textcolor{black}{solid lines} indicate $X_0$ power law fitting: $I_{XX^-} \propto I_{X_0}^{\alpha}$.
	(b) $\alpha$ plotted as a function of $V_g$, indicating a sharp decrease of $\alpha$ with $|V_g|$.}\label{fig:gate_powerlaw}
\end{figure}

Finally, we show that the $XX^-$ kinetics can further be modulated by a gate voltage as well, as predicted from equation \ref{eq:expl}. In \textcolor{black}{Figure \ref{fig:gate_powerlaw}a}, for every $V_g$ value, the incident laser power ($P$) is varied, and the PL intensity of charged biexciton ($I_{XX^-}$) is fitted with exciton intensity ($I_{X_0}$) using the approximate relation $I_{XX^-} \propto I_{X_0}^{\alpha}$ lines (fitting with $I_{XX^-} = P^{\alpha}$ lines is shown in \textbf{Supporting information \textcolor{black}{Figure 9}}). Variation of $\alpha$ exponents as a function of $V_g$ is shown in Figure \ref{fig:gate_powerlaw}b. We note that the maximum occurs near $V_g = 0$. \textcolor{black}{Such a strong modulation of the $XX^-$ kinetics can be understood by using equation \ref{eq:expl}. An increase in $|V_g|$, in turn, enhances the screening in the sample. This affects the stability of $XX^-$ which is of large Bohr radius and relatively weak binding energy. This also triggers an easier dissociation of the charged biexciton ($XX^-$). Hence, the parameter $\gamma_2$ increases driving the power-law exponent towards unity.}

In conclusion, we showed that strong non-equilibrium explains the deviation of the charged biexciton intensity from a square-law dependence on the exciton intensity. The nature of dependence can be modulated by external stimuli, such as gate voltage and temperature. Unlike the neutral exciton, the existence of near-parallel bands between excited and final states for the biexciton removes the usual ``light-cone" restriction during direct radiative recombination. Consequently, the binding energy of the five-particle species is shown to be less than its peak separation from the neutral exciton. The temperature-dependent enhancement of the decay rate of the charged biexciton further supports the rate equation model. The findings elucidate a deeper understanding of biexciton kinetics in monolayer semiconductors.

\section*{Methods:}
 \textbf{Fabrication:} Au/Ag lines are defined by optical lithography using a
 360 nm UV source and AZ5214E resist spin-coated on a Si/SiO2
 substrate with 285 nm thick oxide formed by dry chlorinated thermal oxidation and forming gas annealing. A 20 nm thick Ni film followed by a 40 nm thick Au/Ag film is deposited via DC magnetron sputtering and lifted off by acetone/isopropyl alcohol rinse to form the bottom contact. For device D1, first, a few-layer hBN is exfoliated from bulk crystals and subsequently transferred to a poly dimethyl-siloxane (PDMS) sheet. Then it is transferred onto lithographically pre-patterned Au electrodes using dry-transfer technique. The thickness of the flake is identified by optical contrast. In the second step, 1L WS\tsub{2} flake was transferred and connected with another Au electrode using few-layer graphene. Then top hBN is transferred onto the stack following the same procedure. The device was annealed after each layer transfer at 80$^{\circ}$C for better adhesion. For device D2, a similar layer-by-layer transfer procedure was followed, except Ag pre-patterned electrodes were used instead of Au ones, and a graphene transfer step was not needed. \\
 \textbf{PL measurement:} The devices are loaded or are wire bonded to a closed-cycle optical top window He cryostat and are illuminated with 532 nm continuous wave laser through $\times 50$ objective lens having a numerical aperture of 0.5. Keithley 2636B is used as a DC voltage source for applying gate voltage ($V_g$) to vary the doping in device D1. The steady-state PL response of devices (D1 and D2) is analyzed using a spectrometer with 1800 lines/mm grating. All the measurements are performed in a vacuum with a pressure $< 10^{-4}$ Torr.  \\
\textbf{TRPL measurement:}  We excite the sample with a 531 nm pulsed laser having full-width-at-half-maximum (FWHM) of 48 ps and a repetition rate of 5 MHz. The sync signal from the laser driver is fed to channel 0 of a TCSPC (PicoHarp 300). The emission from the sample is fed to a single photon detector (Micro Photon Devices), the output of which is connected to channel 1 of the TCSPC. The instrument response function (IRF) shows an FWHM of 52 ps and a decay timescale of 23 ps. Using deconvolution, we can accurately estimate down to 10\% of the IRF width \cite{becker2005advanced}. To filter out $XX^-$ luminescence only, we use a bandpass filter (FWHM of 10 nm) centered at 610 nm. \textcolor{black}{For \textit{in situ} steady-state PL spectra, a 50:50 beam splitter is used to divert part of the emission to a spectrometer.}

\section*{ACKNOWLEDGMENTS:}
 Growth of hexagonal boron nitride crystals was supported by the Elemental Strategy Initiative conducted by the MEXT, Japan, Grant Number JPMXP0112101001,  JSPS KAKENHI Grant Number JP20H00354 and JP19H05790. K.M. acknowledges the support from a grant from Science and Engineering Research Board (SERB) under Core Research Grant, a grant from the Indian Space Research Organization (ISRO), a grant from MHRD under STARS, and support from MHRD, MeitY, and DST Nano Mission through NNetRA.
\section*{Competing Interests}
The authors declare no competing financial or non-financial interests.
\section*{Data Availability}
Data available on reasonable request from the corresponding author.

\bibliographystyle{unsrt}

\bibliography{references}
\newpage
\newpage
\newpage
\newpage
\newpage


\section*{Supplementary material}

## Supporting Information:

**1. Derivation of temperature dependent dissociation function.**

With increasing temperature the total decay rate (R), for charged biexciton $(XX^-)$, can be written as

$$R = \tau_{XX^-}^{-1} + k_0 e^{\frac{-Eb}{k_B T}} \tag{1}$$

Where $E_b$ and $\tau_{XX^-}$ are the binding energy and radiative lifetime of $XX^-$ respectively. The Quantum efficiency $(\eta(T))$ for radiative decay is:

$$\eta(T) = \frac{\tau_{XX^-}^{-1}}{R} \tag{2}$$

Equation 2 can be simplified to:

$$\eta(T) = \frac{1}{1 + \kappa e^{-E_b/k_B T}} \tag{3}$$

where $\kappa = k_0 \tau_{XX^-}$ and $k_0$ is a fitting parameter [1, 2]. Since by measuring PL signal we can only probe radiative decay, equation 3 can be fitted with temerature dependent PL intensity variation of $XX^-$, shown in the main text Figure 3b.

**2. Derivation of steady state coupled exciton $(n_{X^0})$ and charged biexciton $(n_{XX^-})$ population equation.**

The rate equations are [3, 4]:

$$dn_{X^0}/dt = -n_{X^0}/\tau_{X^0} + \beta_1 P + \gamma_2 n_f n_{XX^-} e^{\frac{-\Delta E_{XX^-}}{k_B T}} - \beta_2 n_{X^0} - \gamma_1 (n_f + 1) n_{X^0} n_{X_D^-} \tag{4}$$

$$dn_{X_D^-}/dt = -n_{X_D^-}/\tau_{X_D^-} + \beta_2 n_{X^0} + \gamma_2 n_f n_{XX^-} e^{\frac{-\Delta E_{XX^-}}{k_B T}} + n_{XX^-}/\tau_{XX^-} - \gamma_1 (n_f + 1) n_{X^0} n_{X_D^-} \tag{5}$$

$$dn_{XX^-}/dt = -n_{XX^-}/\tau_{XX^-} + \gamma_1 (n_f + 1) n_{X^0} n_{X_D^-} - \gamma_2 n_f n_{XX^-} e^{\frac{-\Delta E_{XX^-}}{k_B T}} \tag{6}$$

at steady sate the differentials with respect to time are zero. We can substitute $n_{XX^-}/\tau_{XX^-}$

from equation 6 to equation 5. Then we get a linear relationship between bright exciton and dark trion.

$$n_{X_D^-} = \tau_{X_D^-} \beta_2 n_{X^0} = A_1 n_{X^0} \tag{7}$$

Now, $n_{X_D^-}$ can be replaced by a function of $n_{X^0}$ in the equation 6 and assuming $n_f \approx n_{XX^-}$:

$$n_{XX^-}/\tau_{XX^-} + \gamma_2 n_{XX^-}^2 e^{-\Delta E/k_B T} = \gamma_1 (n_f + 1) A_1 n_{X^0}^2 \tag{8}$$

From equation 4 we can substitute the $n_{X^0}$ as a function of $n_{XX^-}$ and relation between $n_{XX^-}$ and laser power (P) can be established.

$$\beta_1^2 P^2 C_1 = n_{XX^-}^2 [\gamma_2 \alpha e^{-\Delta E_{XX^-}/k_B T}(B_1)^2 + C_1/\tau_{XX^-}^2] +$$
$$2n_{XX^-}^{3/2} C_1^{1/2} B_1 [1/\tau_{XX^-} + \gamma_2 n_{XX^-} \alpha e^{-\Delta E_{XX^-}/k_B T}]^{1/2} + n_{XX^-}/\tau_{XX^-}(B_1)^2 \tag{9}$$

where $B_1 = (\beta_1 + 1/\tau_x)$ and $C_1 = A_1 \gamma_1 (n_f + 1)$

**Supporting Figures:**

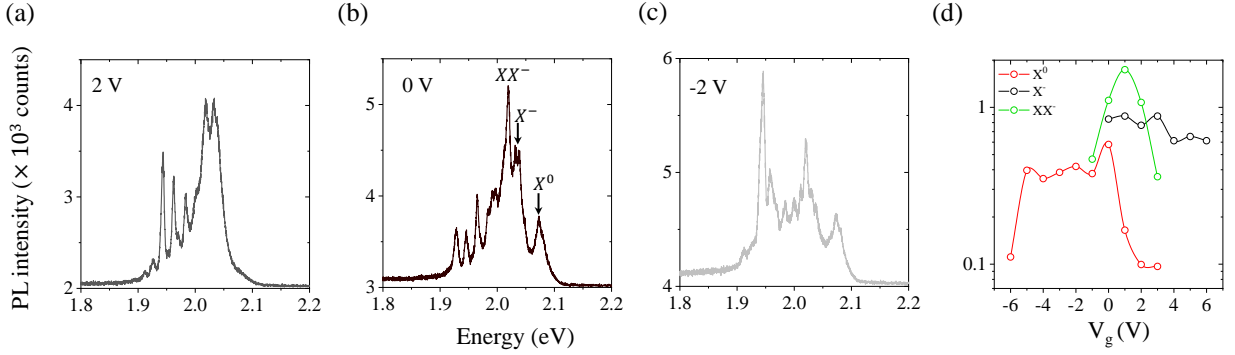

Figure 1: **Gate voltage ($V_g$) dependent PL spectra obtained at 4 K using cw 532nm laser.** The corresponding gate voltages are shown in left side of the spectra, (b) at $V_g = 0$ V signature of $XX^-$ and other PL peaks such as: exciton ($X^0$), trion ($X^-$) are clearly visible. However due to nonpresence of exciton and trion in (a) electron (positive $V_g$) and (c) neutral (negative $V_g$) doping region respectively, formation of $XX^-$ is hindered. (d) Fitted PL peak intensities of all the species noted in main text are plotted with $V_g$ (in log scale). Clearly, after $V_g > 3$ V both $X^0$, and $XX^-$ are not visible.

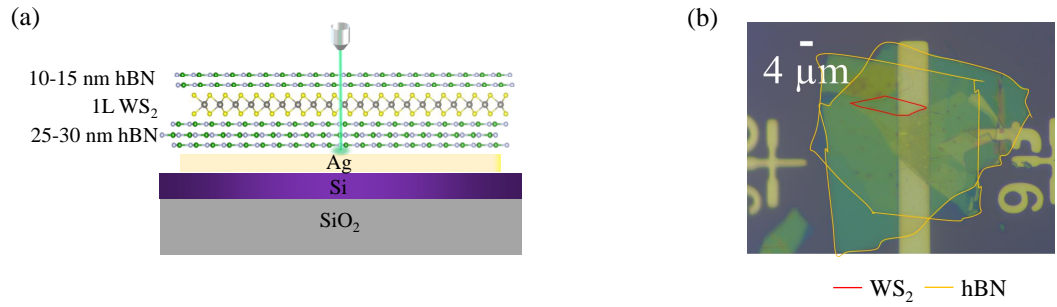

Figure 2: **Schematic and optical image of device D2** (a) Schematic (b) Optical image of Device D2 where top, bottom hBN and 1L WS$_2$ layers are marked with colored borders. Here 1L WS$_2$ is kept floating without any applied gate potential ($V_g$).

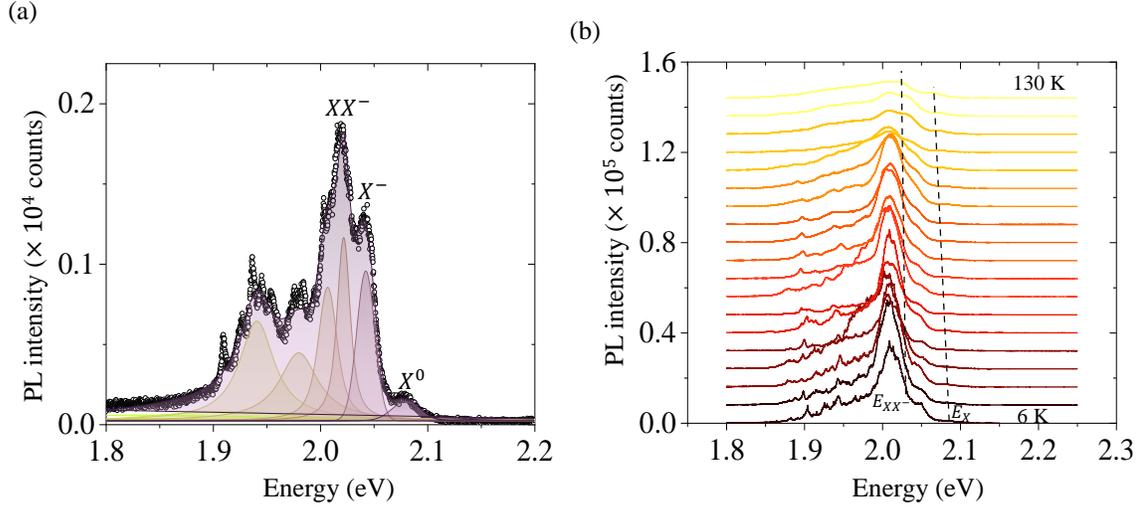

Figure 3: **Fitting of PL spectrum to obtain charged-biexciton ($E_{XX^-}$) peak position and intensity and spectra at varying temperature.** (a) We fit the obtained spectra with 532 nm cw laser excitation to obtain the $XX^-$ peak position ($\approx 2.02$ eV) and intensity. (b) Temperature dependent spectra showing $XX^-$ and $X^0$, intensity values of $XX^-$ peak obtained from this spectra is shown in the main text, Figure 2c.

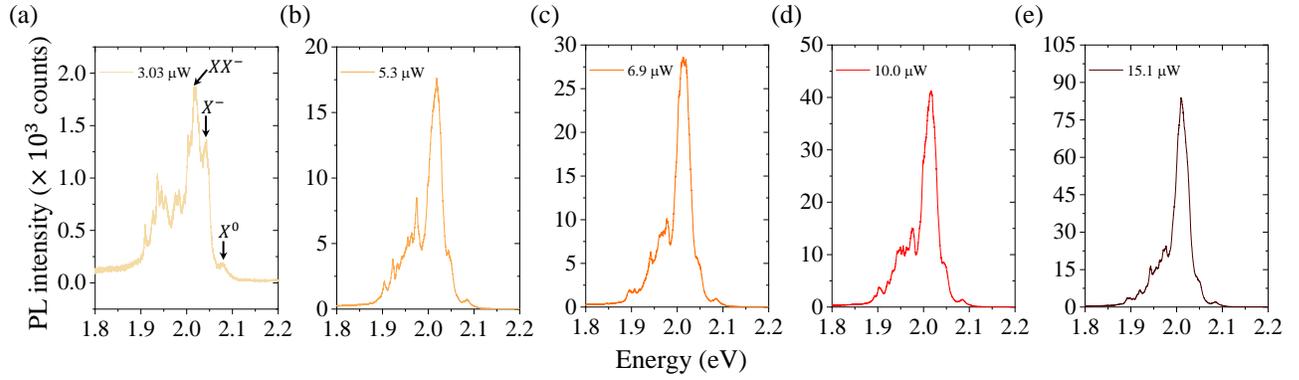

Figure 4: **PL spectra at 4K with varying laser power.** We vary the cw 532 nm laser power from 3.03 $\mu W$ to 15.1 $\mu W$ (from (a) to (e)) at 4K to obtain $XX^-$ intensity variation log-log plot shown in the main text Figure 3b and c.

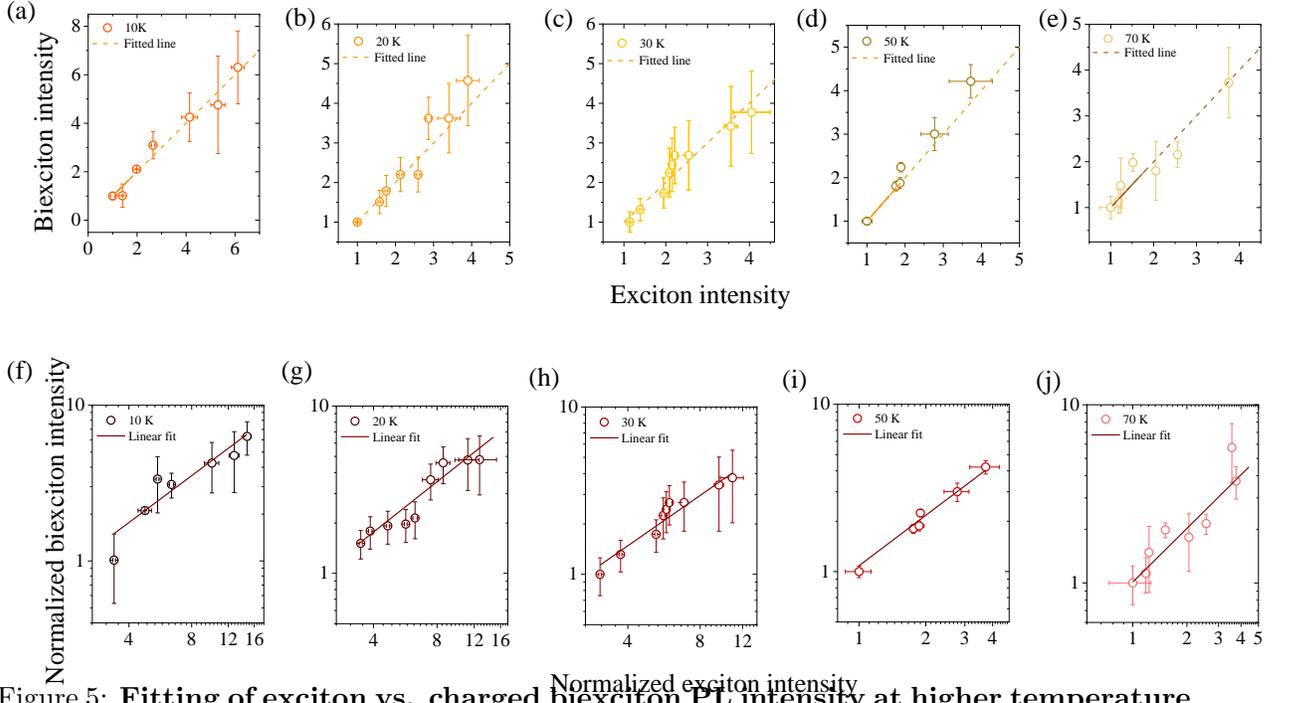

Figure 5: **Fitting of exciton vs. charged biexciton PL intensity at higher temperature points.** Fitting with equation 9 of main text, using dashed line, (top panel, (a)-(e)) and $I_{XX^-} \propto I_{X^0}^\alpha$ (bottom panel, f-j), using solid line, for higher temperature range (T = 10, 20, 30, 50 and 70 K) is shown here. After $\approx$ 10K near unity exciton power law is observed.

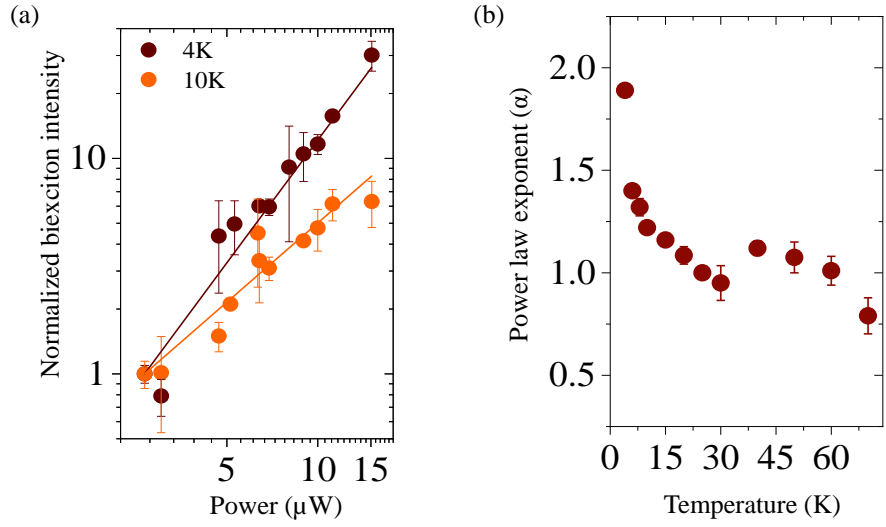

Figure 6: **Charged biexciton intensity with varying laser power at different temperature.** (a) Experimentally obtained $XX^-$ intensity (colored symbols) at 4 and 10K are fitted with $I_{XX^-} \propto P^\alpha$ (solid lines). (b) Fitted power law exponent ($\alpha$) is plotted with varying temperature, showing a downward trend towards near unity power law after $\approx 20$K.

9

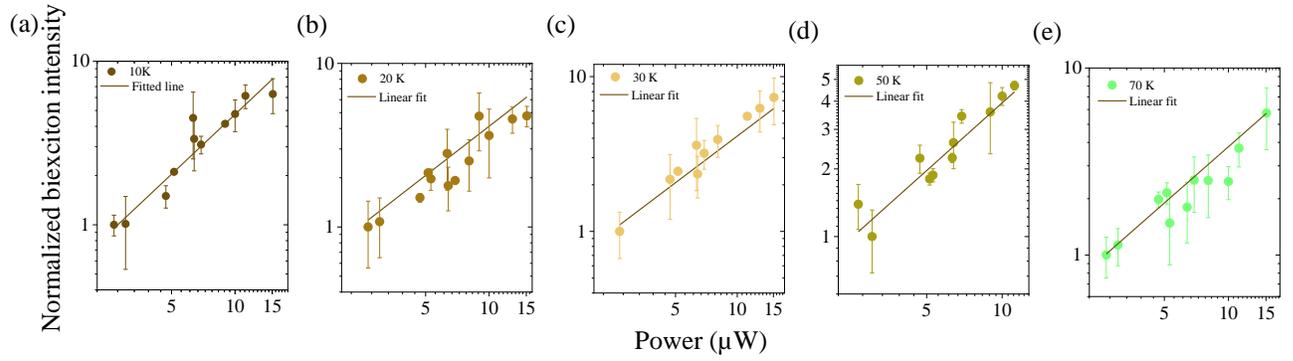

Figure 7: **Laser power law fitting at high temperature points.** Experimentally obtained $XX^-$ PL intensity with laser power at 10, 20, 30, 50 and 70 K ((a)-(e)) are fitted with $I_{XX^-} \propto P^\alpha$ (solid lines). After $\approx$ 20K near unity power law is observed.

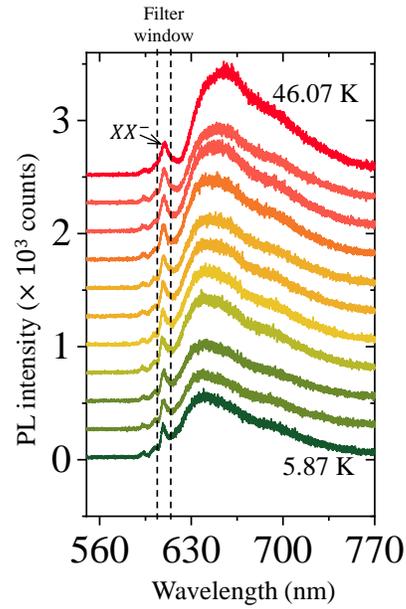

Figure 8: **Temperature dependent <u>in situ</u> PL spectra taken along with TRPL data.** Plot of <u>in situ</u> PL spectra, measured at same temperature points as shown in main text, Figure 4b. The filter window, shown by dashed line, contains $XX^-$ PL emission mostly (marked by arrow).

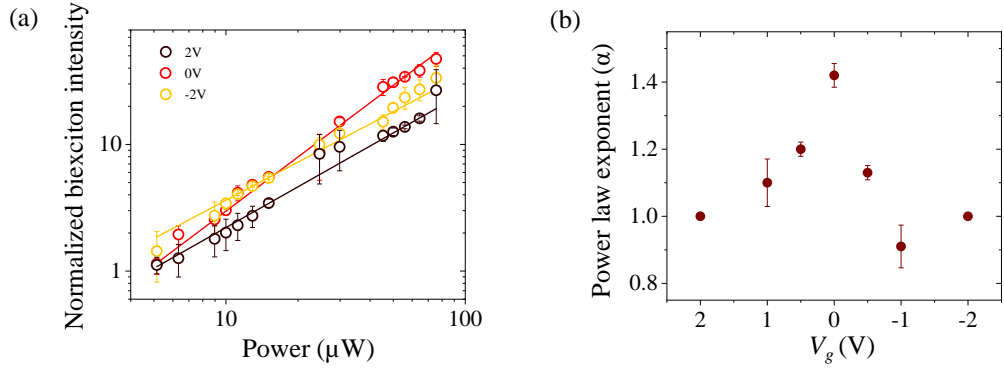

Figure 9: **Gate voltage ($V_g$) dependent kinetics for charged biexciton.** (a) Plot of $XX^-$ peak intensity with varying laser power (P) at different $V_g$. The solid traces indicate power law fitting: $I_{XX^-} \propto P^\alpha$. (b) $\alpha$ plotted as a function of $V_g$, indicating a sharp decrease of it with $|V_g|$.

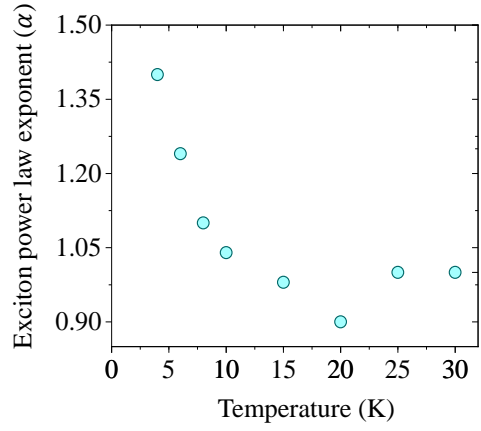

Figure 10: **Exciton power law measurement in sample D1.** With varying Laser power obtained $XX^-$ peak intensity is fitted using $I_{XX^-} \propto I_{X^0}^{\alpha}$ lines and extracted $\alpha$ values are plotted here. This plot is in excellent agreement with main text, Figure 3c.


\end{document}